\documentclass[
 amsmath,amssymb,
 aps,
 prl,
 longbibliography,
 lengthcheck,%
]{revtex4-1}

\usepackage[dvipdfm]{graphicx}% Include figure files
\usepackage{dcolumn}% Align table columns on decimal point
\usepackage{bm}% bold math
\begin{document}

%\preprint{}

\title{Non-universal Efimov Atom-Dimer Resonances in a Three-Component Mixture of $^{6}$Li}


\author{Shuta Nakajima$^{1*}$, Munekazu Horikoshi$^{2}$, Takashi Mukaiyama$^{2,3}$, Pascal Naidon$^{2}$}
\author{Masahito Ueda$^{1,2}$}
\email[]{shuta@cat.phys.s.u-tokyo.ac.jp}
%\thanks{}
\affiliation{$^1$\mbox{Department of Physics, University of Tokyo, 7-3-1, Hongo, Bunkyo-ku, Tokyo 113-0033, Japan}\\
$^2$\mbox{ERATO Macroscopic Quantum Control Project, JST, 2-11-16 Yayoi, Bunkyo-ku, Tokyo 113-8656, Japan}\\
$^3$Center for Frontier Science and Engineering and Institute for Laser 
Science, University of Electro-Communications, 1-5-1 Chofugaoka, Chofu, Tokyo 182-8585, Japan.}


\date{\today}

\begin{abstract}
We observed an enhanced atom-dimer relaxation due to the existence of Efimov states in a three-component mixture of $^{6}$Li atoms. We measured the magnetic-field dependence of the atom-dimer loss coefficient in the mixture of atoms in state $|1 \rangle$ and dimers formed in states $|2 \rangle$ and $|3 \rangle$, and found two peaks corresponding to the degeneracy points of the $|23 \rangle$ dimer energy level and energy levels of Efimov trimers. 
We found that the locations of these peaks disagree with universal theory predictions, in a way that cannot be explained by non-universal two-body properties. We constructed theoretical models that characterize the non-universal three-body physics of three-component $^{6}$Li atoms in the low energy domain.\end{abstract}

%\keywords{}

\maketitle

% \section{Introduction}

% Since the first evidence of the Efimov state\cite{Efimov} in an ultracold Cs atomic gas\cite{Kraemer}, studying few body physics using ultracold atoms has attracted growing interest among researchers in this field. Because we can descrive the systems with only one scattering length,
% identical boson systems have been investigated theoreticaly\cite{} and experimentally\cite{}.
% For example, the universal scaling law\cite{} for the positions of three-body loss  peaks and dips are confirmed \cite{} and the enhancement of atom-dimer loss due to the existence of Efimov states was also observed in Cs system\cite{}. Although the loss enhancement and the recombination minima in a three-body process are explained to some extent by the universal theory, their positions are still shifted from the predictions by universal theories. Lots of efforts have been made to explain such resonance shift in an identical bosons system by taking into account a finite-range corrections[] or magnetic field dependence of the three-body parameter[], it is still an open question how accurately those corrections describe the Efimov spectrum obtained in the experiments.

Since the first experimental evidence of Efimov states \cite{Efimov} in an ultracold cesium atomic gas \cite{Kraemer}, 
few-body physics in ultracold atoms has attracted growing interest. The observation of three-body and atom-dimer loss peaks and dips confirmed very general properties of few-boson systems near unitarity such as the universal scaling laws associated with the existence of Efimov states \cite{Kraemer,Zaccanti,Gross,Pollack,Knoop,Barontini}.
Although these loss enhancements and recombination minima are qualitatively explained by the universal theory, their positions are shifted from universal predictions. While some efforts have been made to explain such resonance shifts by taking into account finite-range corrections \cite{key-5, key-4, Jona-Lasinio}, it is still an open question how accurately those corrections reproduce the observed Efimov spectra.

\begin{figure}[tbp]
\includegraphics[scale=0.455]{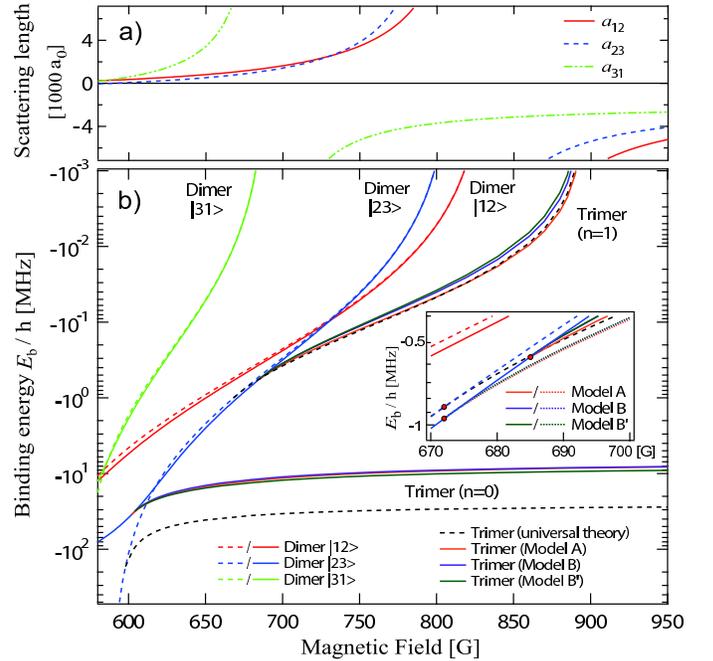}
\caption{\label{fig:Eb} (a) Two-body s-wave scattering lengths in $^6$Li for states $|1\rangle$, $|2\rangle$, and $|3\rangle$ \cite{Bartenstein}. (b) Binding energies $E_{\rm b}$ of the dimer and trimer states in three-component mixture of $^6$Li.
% $|12\rangle$, $|23\rangle$ and $|31\rangle$ are dimer states associated with the Feshbach resonances for $a_{12}$, $a_{23}$ and $a_{31}$, respectively.
Dashed curves represent the energies from universal theory. 
Solid curves for dimers are obtained from two-body coupled-channel calculations. 
Solid curves for trimers are obtained from our non-universal models adjusted to the experimental data.
The inset shows the dimer-trimer crossing near 685~G. 
Dotted lines correspond to non-universal models without adjustment of the three-body parameter.
% Solid curves show the energies from universal theory and dashed curves are calculated from theories include non-universal corrections.
% The dashed lines ($n=0$ and $n=1$) are the binding energies of the
% ground and an excited Efimov trimer states predicted from universal Efimov parameter $a_*$
% obtained by the three-body loss measurement\cite{}.
% Enhancement of the atom-dimer loss occurs when Efimov trimer states cross the binding energy of $|2\rangle|3\rangle$ dimer (arrows).
}
\end{figure}

Recently, it has turned out that a three-component Fermi gas of ${}^6$Li offers another intriguing system to investigate universal few-body physics. Ottenstein \textit{et al.}\cite{Ottenstein} and Huckans \textit{et al.}\cite{Huckans} observed the enhancement of the three-body loss
at 130~G and 500~G in the mixture of fermionic $^6$Li atoms in the three lowest-energy hyperfine states $|F;m_F\rangle= |1/2;1/2\rangle$, $|1/2;-1/2\rangle$ and $|3/2;-3/2\rangle$, which we label as $|1\rangle$, $|2\rangle$ and $|3\rangle$ respectively. Another resonance was later observed at 895~G \cite{Williams}.
It was argued that those three-body loss enhancements are due to the existence of Efimov states \cite{Floerchinger,Braaten_Li,Naidon,Wenz,Braaten2}, 
as in the case of identical bosons.
In such a system, however, because of their fermionic nature, only distinguishable particles can interact via s-wave scattering.
Thus interactions for the respective combinations are described
by three different two-body scattering lengths $a_{12}$, $a_{23}$ and $a_{31}$, which diverge at 834~G, 811~G, and 690~G respectively due to Feshbach resonances (see Fig. \ref{fig:Eb}(a)).
For each of these resonances there is a weakly bound dimer state
% with binding energy $E_{\rm b}=\hbar^2/2ma_{ij}^2$
which we designate as $|12\rangle$, $|23\rangle$ and $|31\rangle$, respectively (see Fig. \ref{fig:Eb}(b)).
Because these Feshbach resonance regimes overlap, the three scattering lengths can be varied simultaneously with a magnetic field, making it possible to realize a strongly-interacting multi-component system and access a whole new quantum phase of matter \cite{Bedaque}. 
For this purpose, an accurate understanding of the low-energy few-body physics is necessary. 
As in the case of identical bosons, this requires an accurate determination of the Efimov spectrum.
% This will allow one to access a whole new quantum phase of matter in the real experiment. Although the complete description of a few-body physics in a multiple-component fermion system based on unversal theories is an even greater challenge, it will improve our understanding of a few-body physics. As is the case of identical bosons, conclusive understanding of the Efimov physics will require more detailed and accurate information on the Efimov spectrum.

%%%%%%%%%%%%%%%%%%%%%%%%%%%%%%%%%%%%%%%%%%%%%%%%%%%%%%%%%%%%
%%%%%%%%%%%%%%%%%%%%%%%%%%%%%%%%%%%%%%%%%%%%%%%%%%%%%%%%%%%%

% These interactions are described
% by three dierent two-body scattering lengths a12, a23
% and a13 for the respective combinations (see g. 1 a)).
% 6Li has the unique advantange that there are broad, overlapping
% Feshbach resonances for all these combinations.
% For each of these resonances there is a weakly bound
% dimer state with binding energy Eij / 1=a2
% ij , which we
% designate j12i, j23i and j13i respectively (see g. 1 b))

In this Letter, we report the observation of two resonantly enhanced atom-dimer loss peaks in an atom-dimer mixture of $^6$Li corresponding to the degeneracy points between the binding energy of $|23\rangle$ dimers and the ground ($n=0$) and excited ($n=1$) Efimov trimer states (see Fig. \ref{fig:Eb}(b)). 
These peaks have been predicted by Braaten \textit{et al.} from theory based on universality \cite{Braaten2}.
We compare the observed peak positions with universal theory predictions, and find significant deviations which cannot be explained by two-body physics only. We then construct non-universal models to interpret them.

In our experiment we used an all-optical method to prepare a degenerate two-component Fermi gas of $^6$Li atoms in the two lowest hyperfine states of $|1\rangle$ and $|2\rangle$ as described in detail in \cite{Inada}. To prepare a mixture of $|1\rangle$ atoms and $|23\rangle$ dimers in equal population, we started with an imbalanced mixture of atoms in state $|1\rangle$ and $|2\rangle$ whose population ratio is $|1\rangle:|2\rangle=2:1$.
Evaporative cooling was performed at the magnetic field of 300~G where the amplitude of the scattering length between $|1\rangle$ and $|2\rangle$ shows a local maximum. The total number of atoms before dimer creation was $\sim 10^6$.
To achieve a very low collision energy, we adiabatically transferred this mixture 
into a larger volume hybrid magnetic/optical trap with smaller oscillation frequencies. 
The trap frequencies for atoms $\omega^{\rm A}$ were experimentally measured and approximately given by $\omega^{\rm A}_x=2\pi \times \sqrt{110^2-0.73B}$ Hz, $\omega^{\rm A}_y=2\pi \times \sqrt{50^2-0.31B}$ Hz and $\omega^{\rm A}_z=2\pi \times \sqrt{7.2^2+0.45B}$ Hz in $x, y$ and $z$ directions respectively, where $B$ is the strength of magnetic field in Gauss.
Since the magnetic moments of $|1\rangle$ atoms and $|23\rangle$ dimers are different, the trap frequencies for dimers $\omega^{\rm D}$ are different from the ones for $|1\rangle$ atoms, especially below 650 G. The trap frequencies for dimers were calculated using the magnetic moment of $|23\rangle$ dimers. Deviation of $\omega^{\rm D}$ from the $\omega^{\rm A}$ for $x, y$ and $z$ directions at 580~G were about $+6$, $+27$, $-26$ \% respectively.

\begin{figure}[tbp]
\includegraphics[scale=0.4]{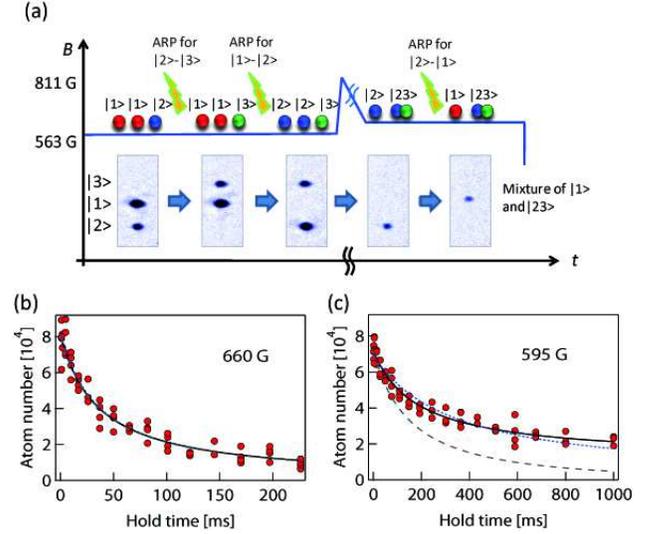}
\caption{\label{fig:sequence} (a) Experimental procedure to prepare atom-dimer mixtures by using a multiple-stage adiabatic rapid passage (see text).
Because the dimers cannot be imaged by resonance light, 
we dissociate the dimers by applying a magnetic field pulse to check the population balance of the atoms and dimers.
(b) and (c): Typical time evolution of the number of atoms in state $|1\rangle$ at 660~G and 595~G. The blue dotted curve shows an analytic fitting function assuming the pure two-body loss and the black solid curve is obtained by a numerical fitting that includes the effect of dimer-dimer loss. Gray dashed curves show time evolution of the numbers of molecules calculated from the rate equations.}
\end{figure}

To create the atom-dimer mixture, we used a multiple-stage adiabatic rapid passage (ARP) as shown in Fig. \ref{fig:sequence}(a).
By applying ARP at 563~G, we first transferred the atoms in state $|2\rangle$ to state $|3\rangle$. Then we transferred the atoms in state $|1\rangle$ to state $|2\rangle$ by ARP. Thus we created an imbalanced mixture of $|2\rangle$ and $|3\rangle$ whose population ratio is $|2\rangle:|3\rangle=2:1$.
%  (See Fig. \ref{fig:sequence}(a)). 
After the preparation of the imbalanced $|2\rangle$ - $|3\rangle$ mixture, 
we quickly ramped the magnetic field up to 811~G, and then swept down to the field of interest (580~G-760~G) in 300~ms to adiabatically create $|23\rangle$ dimers. At this point, excess atoms in state $|2\rangle$ were still left in the trap and the last stage of ARP was applied to transfer the atoms in state $|2\rangle$ to state $|1\rangle$ to obtain a $|1\rangle$ atom - $|23\rangle$ dimer mixture.

%%%%%%%%%%%%%%%%%%%%%%%%%%%%%%%%%%%%%%%%%%%%%%%%%%%%%%%%
%%%%%%%%%%%%%%%%%%%%%%%%%%%%%%%%%%%%%%%%%%%%%%%%%%%%%%%%

To study the decay of an atom-dimer mixture, we measured the remaining fraction of atoms in $|1\rangle$ after a holding time of $1\sim1000$ ms. The magnetic-field dependence of the atom-dimer loss rate was measured by repeating this sequence with various magnetic field values. Therefore the last stage of ARP was applied at a different magnetic field, and the condition for perfect ARP was checked at each magnetic field. When we take the absorption images, we pulsed on the magnetic field gradient and spatially separated each spin component. This allowed us to check that there was no excess atoms in states $|2\rangle$ and $|3\rangle$ during the loss measurements. Although the temperature of the atom-dimer mixture is very low ($\sim$100~nK), it is close to the Fermi temperature due to the very low density. Therefore, we assume that the density distributions of atoms and dimers are Gaussian.
Then, the number of atoms $N_{\rm A}(t)$ and that of dimers $N_{\rm D}(t)$ evolve in time according to the following coupled rate equations,
\begin{eqnarray}
\dot{N}_{\rm A}=-\Gamma N_{\rm A} &-&C \beta \frac{N_{\rm D}}{V_{\rm D}} N_{\rm A}, \\
\dot{N}_{\rm D}=-\Gamma N_{\rm D} &-&C \beta \frac{N_{\rm A}}{V_{\rm A}} N_{\rm D}-\alpha \frac{N_{\rm D}}{V_{\rm D}} N_{\rm D},
\end{eqnarray}
where $\beta$, $\alpha$ and $\Gamma^{-1}=10$ s are the atom-dimer loss coefficient, the dimer-dimer loss coefficient and the one-body loss rate, respectively.
Here,
% $\bar{n}_{\rm A}=N_{\rm A}/(2\pi)^{3/2}\sigma_{x}^{\rm A} \sigma_{y}^{\rm A} \sigma_{z}^{\rm A}=[m\bar{\omega_{\rm A}}^2/4\pi k_BT]^{3/2}N_{\rm A}$ and 
% $\bar{n}_{\rm D}=N_{\rm D}/(2\pi)^{3/2}\sigma_{x}^{\rm D} \sigma_{y}^{\rm D} \sigma_{z}^{\rm D}=[m\bar{\omega_{\rm D}}^2/2\pi k_BT]^{3/2}N_{\rm D}$
$V_{\rm A}=\sqrt{8}\pi^\frac{3}{2}\sigma_{x}^{\rm A} \sigma_{y}^{\rm A} \sigma_{z}^{\rm A}$ and 
$V_{\rm D}=\sqrt{8}\pi^\frac{3}{2}\sigma_{x}^{\rm D} \sigma_{y}^{\rm D} \sigma_{z}^{\rm D} = V_{\rm A} (\bar{\omega}^{\rm A} / \sqrt{2}\bar{\omega}^{\rm D})^3$
are the effective atomic and molecular volumes, where
$\sigma_{x,y,z}^{\rm A}=\sqrt{k_BT/m({\omega_{x,y,z}^{\rm A}})^2}$ and
$\sigma_{x,y,z}^{\rm D}=\sqrt{k_BT/2m({\omega_{x,y,z}^{\rm D}})^2}$ are the atomic and dimer cloud widths.
We denote the mass of $^6$Li by $m$, and $\bar{\omega}^{\rm A}$ and $\bar{\omega}^{\rm D}$ are the geometric means of the trap frequencies for atoms and dimers.
$C=\prod_{i} \sqrt{4/(2+(\omega_i^{\rm A}/\omega_i^{\rm D})^2)}\simeq 8/\sqrt{27}$ is a numerical constant which results from the difference of the density distributions of the atoms and dimers due to unequal masses \cite{Knoop}. 

%Since the decay time scale of atom-dimer loss above 650 G is far shorter than the time scale of dimer-dimer and one-body loss, a simple decay curve, $N(t)=N_0/(1+\gamma N_0 t)$, can be used as a fitting function which only includes atom-dimer loss. Here, $N(t)=N_{\rm A}(t)=N_{\rm D}(t)$, $N(0)=N_{\rm A}(0)=N_{\rm D}(0)$ and $\gamma=\sqrt{8/27}(m\bar{\omega}^2/2\pi k_B T)^{3/2}$.
%Below 650 G, the dimer-dimer loss and one-body loss need to be taken into account and we solved the rate equation with those terms.

We determined the atom-dimer loss coefficient $\beta$ by fitting our data with the solution of the rate equations using $\Gamma$ and $\alpha$ which were experimentally determined in advance. We checked that the cloud size remains approximately constant; therefore we assume the constant volume and
ignore heating effect in our analysis. 
We found that the decay time scale of atom-dimer loss above 650~G is far shorter than the time scales of dimer-dimer and one-body loss and the decay curve can be described using two-body loss fitting. However, below 650~G, the dimer-dimer loss and one-body loss need to be taken into account (see Fig. \ref{fig:sequence}(b) and (c)).

%%%%%%%%%%%%%%%%%%%%%%%%%%%%%%%%%%%%%%%%%%%%%%%%%%%%%%%%%%%%
\begin{figure}[tbp]
\includegraphics[scale=0.52]{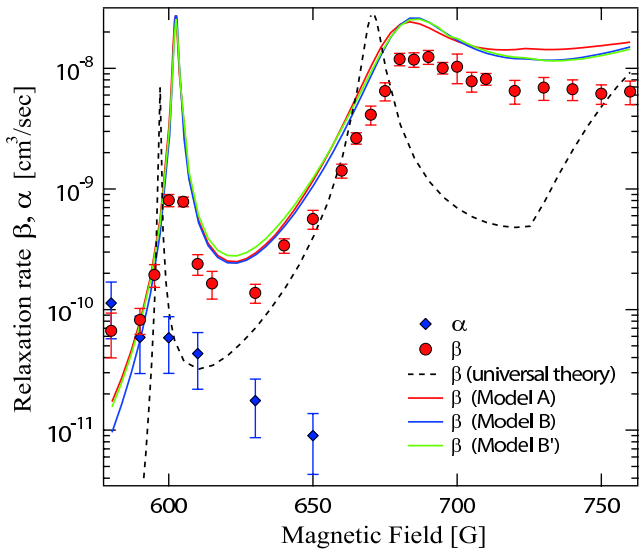}
\caption{\label{fig:ADL} Magnetic-field dependence of the atom-dimer loss coefficient $\beta$ (red circles) and dimer-dimer loss coefficient $\alpha$
(blue diamonds) in the mixture of atoms in state $|1\rangle$ and dimers of $|23\rangle$.  
The black dashed curve is the calculated $\beta$ from universal theory. Red, blue, and green solid curves show the $\beta$ values calculated from our non-universal models A, B and B' respectively. }
\end{figure}
%%%%%%%%%%%%%%%%%%%%%%%%%%%%%%%%%%%%%%%%%%%%%%%%%%%%%%%%%%%%

Figure \ref{fig:ADL} shows the measurement of the atom-dimer loss coefficient $\beta$ (red circles) and the dimer-dimer loss coefficient $\alpha$ (blue diamonds) for the $|23\rangle$ dimers, as a function of magnetic field. 
The error bars for $\alpha$ show the systematic error of our measurements arising from the uncertainty in the number of atoms and size measurements.
The error bars for $\beta$ arise from the statistical error of size mesurements and error propagation of $\alpha$.
The systematic error arising from the uncertainty in the number of atoms, which is estimated to be $\pm$22\% and may shift the data linearly, is not
included in these error bars.
We observe two maxima of the atom-dimer loss coefficient near 602~G and 685~G.

%The error bars for $\alpha$ come from our systematic errors for atom number and trap size ditermination.
%The lower limit of $\beta$ estimeted from upper value of the $\alpha$ and upper limit from lower value of the $\alpha$.
%$\beta$ does not include our systematic error like abovearises from our uncertainty of the absolute atom number ($\pm$22\%), 
%which may shift the result vertically on the graph.

The atom-dimer loss is expected to increase at the magnetic field where the energy level of an Efimov state intersects with the energy level of the $|23\rangle$ dimer \cite{Braaten2}. Since there is no other mechanism that causes an enhanced atom-dimer loss, the two peaks observed in Fig. \ref{fig:ADL} should correspond to such Efimov resonances. Because the lower peak is expected to be associated with a ``ground-state" Efimov trimer, our finding provides the first experimental evidence for this ``ground-state" Efimov trimer in the three-component mixture of $^6$Li.

We now analyze the data. The usual universal theory \cite{Braaten_Li, Naidon, Braaten2} for this
kind of three-body systems relies only on the 3 scattering lengths to
describe the two-body interactions and a short-range three-body parameter
$\Lambda e^{i\eta}$, where $\Lambda$ fixes the phase of the three-body
wave function at short distance and $\eta>0$ phenomenologically models
losses occuring when 3 atoms come close. These free parameters were
previously determined to be $\Lambda=0.885\; a_{0}^{-1}$ and $\eta=0.016$
from the zero-energy resonance at 895~G lying right in the universal
region of large scattering lengths. Figures. \ref{fig:Eb}(b) and \ref{fig:ADL} show the universal
predictions for the trimer energy and atom-dimer loss coefficient using these
previously determined parameters. While they qualitatively explain 
the presence of the two resonances, there is significant disagreement.
This is not surprising because the resonances investigated here are
significantly away from the universal region. Near the resonance at 602~G, the scattering lengths are on the order of the range of
the interactions
(the van der Waals length $\ell_{vdW} \sim 60~a_0$)
and the binding energy of $|23\rangle$ dimers differs by 57 \% from the universal form $-\hbar^{2}/ma_{23}^{2}$.
Near the resonance at 685~G, the scattering lengths are much larger,
yet the binding energy still differs from the universal form by 8 \%. If we follow the predicted universal trimer
energy, this deviation suggests that the resonance point
should be shifted to 685~G, which is interestingly very close to the
observed peak location.

%
% \begin{figure}
% \includegraphics[scale=0.6]{LossRatePlot.eps}
% \caption{\label{fig:theory}(color online) Atom-dimer loss rate. Dots: experimental data. Dotted line: universal
% theory. Red line: model A. Green line: model B and B$^{\prime}$ (dashed). }
% \end{figure}

To check this point and make a consistent analysis of our data, the minimal requirement is to design a theory which accurately describes the non-universal two-body
physics, in particular the dimer energies.
The most straightforward way to achieve this is to solve the three-body
problem with zero-range interactions parametrized by energy-dependent scattering lengths. This can be easily implemented using the Skorniakov - Ter-Martirosian coupled integral equations \cite{Skorniakov, key-6, Braaten2}. The scattering length $a(k)$ should accurately reproduce the asymptotic two-body physics
at positive and negative energies $E=\frac{\hbar^{2}k^{2}}{m}$. In
particular, the dimer binding energy $E_{\rm b}=-\frac{\hbar^{2}\kappa^{2}}{m}$
is given by $1/a(i\kappa)=\kappa$ . The universal limit is retrieved
for an energy-independent scattering length $1/a(k)=1/a$. A realistic
analytical expression near a Feshbach resonance can be derived from a two-channel
model, for instance with separable Gaussian interactions \cite{Jona-Lasinio}. It leads
to\[
\frac{1}{a(ik)}=e^{-(bk)^{2}}\frac{1}{a_{bg}}\left(1-\frac{1}{(k\sigma(B))^{2}+\frac{B-B_{0}}{\Delta B(B)}}\right)^{-1}+f(k)\]
where the first term contains the resonance
parameters $a_{bg}$, $B_{0}$, $\Delta(B)$, and $\sigma(B)$ which
are fitted to reproduce accurately the zero-energy scattering length,
the effective range, and the last dimer binding energy - all obtained
from a two-body coupled-channel calculation for each hyperfine configuration.
The high-energy term $f(k)$ is found to be $f_{\rm A}(k)=k \mbox{Erf} (kb)$ which
behaves as $k$ at large $k$. While the high-energy (short-distance)
behaviour is irrelevant for the low-energy two-body physics, it does
change the three-body phase at short distance. It is known however that
two-body physics only cannot determine that phase in general~\cite{DIncao2}. 
Although we could adjust the high-energy two-body form to effectively set the 3-body phase \cite{Jona-Lasinio},
we rely for that purpose on the three-body parameter $\Lambda e^{i\eta}$,
which appears as an upper bound of the integral in the STM equations. To
check that our theory does not depend on the two-body high-energy
form, we consider an alternative form $f_{\rm B}(k)=\frac{bk^{2}}{1+(bk)^{2}}$
which tends to a constant $\frac{1}{b}$ at large $k$. In all cases,
we choose a form which is independent of the low-energy parameters,
and is therefore magnetic-field independent.

Using these two models, which we refer to A and B, we first adjust
the three-body parameters $\Lambda$ and $\eta$ to fit the previously measured loss coefficient near the zero-energy
resonance at 895~G. As in the usual Efimov theory,
the choice of $\Lambda$ is not unique and for completeness we consider
two values for model B, which we refer to B and B'.
We found that all models yield exactly the same universal resonant loss profile near 895~G. Using those
parameters, 
% we then calculate the $|1\rangle-|23\rangle$ atom-dimer loss rate coefficient - see Fig.\ref{fig:ADL}. 
we then investigate the atom-dimer properties.
All models give essentially the same prediction for the second resonance near at 685~G, but different ones for the first resonance at 602~G, probably
reflecting its strongly non-universal character. The predicted location
of the second resonance is around 672~G (see inset of Fig. \ref{fig:Eb}), which is not shifted from
the universal prediction as we naively expected - we also
checked that the same result is obtained with a simple effective range
model. 
We conclude that our experimental data show that the short-range
physics parametrization cannot be constant altogether. 
Therefore, if we keep our two-body parametrization, the three-body parameter $\Lambda$ must depend on energy, and possibly magnetic field. This seems reasonable, since the two-body parameters $a(k)$ are already required to be energy-dependent in order to describe the non-universal two-body physics.

%
% \begin{table}
% \begin{tabular}{|c|c|c|c|c|}
% \hline 
%  & 602 G ($E=?$) & 685 G ($E=?$) & 810 G ($E=0$) & 895 G ($E=0$)\tabularnewline
% \hline
% \hline 
% Model A & (0.0405, 0.008) & (0.0460, 0.08) & (0.0484,-) & (0.0484, 0.033)\tabularnewline
% \hline 
% Model B & (0.0111, 0.020) & (0.0152, 0.20) & (0.0186,-) & (0.0186, 0.011)\tabularnewline
% \hline 
% Model B$^{\prime}$ & (0.141, 0.035) & (0.185, 0.30) & (0.244,-) & (0.244, 0.016)\tabularnewline
% \hline
% \end{tabular}
\begin{table}[tbp]
\caption{\label{tbl:1} Fitted three-body parameters $\Lambda$ and $\eta$ for different magnetic
fields (energies) corresponding to different resonant points. 
These values are obtained for $b=38.85~a_0$ in Model A, and $b=43.93~a_0$ in Model B and B',
which best reproduce the effective ranges of the interactions.
We assume that the values for $\Lambda$ are the same at zero energy.}
\begin{tabular}{|c|c|c|c|c|}
\hline 
  Model &  602~G  &  685~G  &  810~G ($E=0$)  &  895~G ($E=0$) \tabularnewline
\hline
\hline 
 {A } &  {$\Lambda$=0.0405 } &  {$\Lambda$=0.0460 } &  {$\Lambda$=0.0484 } &  {$\Lambda$=0.0484}\tabularnewline
\hline 
 {B } &  {$\Lambda$=0.0111 } &  {$\Lambda$=0.0152} &  {$\Lambda$=0.0186 } &  {$\Lambda$=0.0186}\tabularnewline
\hline 
 {B$^{\prime}$ } &  {$\Lambda$=0.141} &  {$\Lambda$=0.185 } &  {$\Lambda$=0.244 } &  {$\Lambda$=0.244}\tabularnewline
\hline
\hline 
 {A } &  {$\eta$=0.008 } &  {$\eta$=0.08 } &  {- } &  {$\eta$=0.033}\tabularnewline
\hline 
 {B } &  {$\eta$=0.020 } &  {$\eta$=0.20 } &  {- } &  {$\eta$=0.011}\tabularnewline
\hline 
 {B$^{\prime}$ } &  {$\eta$=0.035 } &  {$\eta$=0.30 } &  {- } &  {$\eta$=0.016}\tabularnewline
\hline
\end{tabular}

\end{table}

We proceed to map out this energy dependence by fitting each resonance
with our models - see Table \ref{tbl:1}. The variation of $\Lambda$ with magnetic
field is smooth and almost linear, but it corresponds to a non linear
dependence on energy. The variation of $\eta$ is not monotonic and varies over one order of magnitude.
This is not totally unexpected since it describes underlying
loss processes which are at present unknown. We then perform a quadratic
fit for both $\Lambda$ and $\ln\eta$ to estimate those parameters
at any magnetic field, and plot the corresponding atom-dimer loss
coefficients for each model, as well as the Efimov trimer energies - see Figs. \ref{fig:Eb}(b) and \ref{fig:ADL}.
The results are essentially the same, indicating
that our analysis is model-independent. It reproduces the experimental
data for $B>$ 610~G up to a factor of 2. The first resonance cannot
be very well reproduced, most probably because it is located at an
energy which lies at the margin of validity of our zero-range calculation.

%%%%%%%%%%%%%%%%%%%%%%%%%%%%%%%%%%%%%%%%%%%%%%%%%%%%%%%%%%%%%%%%%%%%%%%%%%%%%%%%%%%%%%%%%

In summary, we have measured the atom-dimer loss coefficient $\beta$ in a mixture of $^{6}$Li hyperfine state $|1\rangle$ and $^{6}$Li dimers in state $|23\rangle$, and found two loss maxima near 602~G and 685~G. We attributed these peaks to the crossings between the atom-dimer threshold and two (ground and excited) Efimov states. 
We found significant deviations from the universal predictions, and showed that they cannot be explained simply by taking into account the non-universal two-body physics. Our work therefore provides evidence for the non-universal character of short-range three-body behaviour, which we quantified by variations of the three-body short-range parameters. Although these parameters are different for different two-body models, they lead to a model-independent interpretation of our data, which predicts the non-universal properties of the two Efimov trimers in three-component $^{6}$Li. Understanding these three-body short-range variations will be a challenging task in the future.

During the preparation of this paper, we became aware of similar results reported by T. Lompe et al. \cite{Lampe2}.

%Becaouse these loss resonances are kinds of "Feshbace resonance" for atoms and dimers, 
%the atom-dimer mixture near the loss resonance may provid a way to measure the binding energy of the Efimov trimer.  

\begin{acknowledgments}
We thank Y. Castin and L. Pricoupenko for useful discussions, and P. S. Julienne and E. Tiesinga for providing accurate lithium interaction potentials.
S.N. acknowledges support from the Japan Society for the Promotion of Science.
\end{acknowledgments}

% \bibliography{basename of .bib file}

%\textbf{Acknowledgements}\\
%The authors acknowledge ... for comments and discussions.

\end{document}